\documentclass[prb,twocolumn,superscriptaddress]{revtex4}

\usepackage{mathtools}
\usepackage{lipsum}
\usepackage{amsmath,amssymb}
\usepackage{MnSymbol}
\usepackage{graphicx}
\usepackage{color}
\usepackage{hyperref}
\usepackage{url}
\usepackage{slashed}
\usepackage{slashed,bbm}
\usepackage{graphics,psfrag,epsfig}
\usepackage{dsfont}
\usepackage{enumerate}
\usepackage{pbox}
\usepackage{braket}
\usepackage[retainorgcmds]{IEEEtrantools}

\DeclareMathSizes{10}{10}{6}{4}

\definecolor{violett}{rgb}{.3,0,.7}

\graphicspath{{./figures/}}

\begin{document}
	
	\title{Three-Dimensional Fractional Topological Insulators in Coupled Rashba Layers}
	
	\author{Yanick Volpez}
	\affiliation{Department of Physics, University of Basel, Klingelbergstrasse 82, CH-4056 Basel, Switzerland}
	
	\author{Daniel Loss}
	\affiliation{Department of Physics, University of Basel, Klingelbergstrasse 82, CH-4056 Basel, Switzerland}
	
	\author{Jelena Klinovaja}
	\affiliation{Department of Physics, University of Basel, Klingelbergstrasse 82, CH-4056 Basel, Switzerland}
\date{\today}	
	
	\begin{abstract}
We propose a model of three-dimensional topological insulators consisting of weakly coupled electron- and hole-gas layers with Rashba spin-orbit interaction stacked along a given axis. We show that in the presence of strong electron-electron interactions the system realizes a fractional strong topological insulator, where the rotational symmetry and condensation energy arguments still allow us to treat the problem  as quasi-one-dimensional with bosonization techniques. 
We also show that if Rashba and Dresselhaus spin-orbit interaction terms are equally strong, by doping the system with magnetic impurities, one can bring it into the Weyl semimetal phase.

\end{abstract}
%
	\maketitle

	\begin{figure}[b]
		\centering
		\includegraphics[width=.40\textwidth]{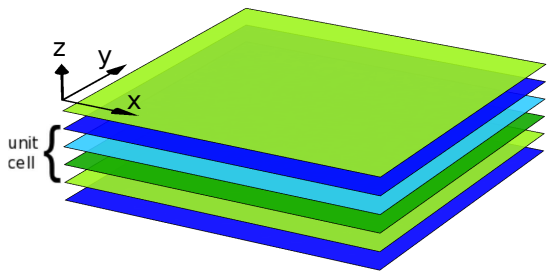}
		\caption{Schematic representation of  the system formed by tunnel-coupled layers with charge carriers. The unit cell consists of two electron (blue) and two hole (green) layers. The color brightness  encodes the two different signs of the SOI. }
		\label{setup}
	\end{figure}

\section{Introduction}
In recent years, the study of topological phases of matter has become one of the prominent subjects in condensed matter research. Soon after the theoretical prediction  and experimental confirmation of topological band insulators (TI) in two (2D)\cite{KaneMele20051,KaneMele20052,Bernevig20061,Bernevig20062,Koenig2007} and three dimensions (3D) \cite{FuKane2007,Moore2007,Hsieh2008,Hsieh2009}, it was theoretically shown that the class of topologically non-trivial matter is much larger and the corresponding phases even more exotic once interacting systems are considered that can allow phases hosting gapless excitations with fractional charges or spin quantum numbers \cite{Levin2009,Maciejko2011,KlinovajaTserkovniak2014}. The realization of such unconventional phases in Nature is not only of fundamental  interest, but also promising for applications such as topological quantum computation, where Fibonacci anyons can serve as qubits which allow for universal quantum computation \cite{Freedman20021}. However, the basic ingredients for obtaining a Fibonacci phase are parafermions, also called fractional Majorana fermions, which emerge only in the presence of electron-electron interactions. Many proposal for experimental realizations of parafermions rely on a combination of superconductivity and fractional TIs \cite{Cheng2012,Lindner2012,Vaezi2013,Clarke2013,KlinovajaLoss2014a,KlinovajaLoss2014b}. So far, fractional TIs still lack experimental realization and it is thus of great importance searching for models possibly realizable in future experiments.

It is the purpose of this Letter to introduce a model which shows how, in principle, a 3D fractional TI can be engineered. We generalize the approach of weakly coupled wires \cite{Kane2002} to 3D by considering a stack of weakly coupled 2DEG layers.  Although the coupled wires approach is a very successful method for theoretically constructing 2D \cite{Poilblanc1987,Gorkov1995,Kane2002,Klinovaja2013,Teo2014,KlinovajaLoss2014c,Meng2014,KlinovajaTserkovniak2014,Neupert2014,Sagi2014,Klinovaja2015,Peter} and 3D \cite{Sahoo2015,Meng2015,SagiOreg2015} topological systems, the coupled layers approach \cite{Luka2016} is simpler to handle and is physically more transparent when describing 3D systems. We consider a stack of 2D layers with Rashba spin-orbit interaction (SOI) weakly tunnel coupled to each other. Such a system could be realized in a semiconductor superlattice where the 2DEGs form at heterojunctions and the SOI can be controlled with electrical gates \cite{Nitta1997,Dettwiler2017}. Alternatively, one could realize our setup in a van der Waals heterostructure, by stacking a carefully chosen sequence of different atomically thin layers on top of each other\cite{Novoselov2016,smet,Novoselov2013,Novoselov2012,Balatsky2013,carlos}. 

The paper is organized as follows. In Section~\ref{Model} we introduce the system composed of weakly coupled layers. In Section~\ref{3DTI}, we study its properties in the non-interacting regime. We derive the bulk spectrum, discuss the computation of the topological invariant, and show the existence of gapless surface states using analytical and numerical methods. We conclude that the non-interacting model realizes a 3D TI. This sets the stage for the main part of the work presented in Section~\ref{FTI} -  the fractional topological phase.
We identify the regime where the interacting system forms a fractional strong 3D TI \cite{Meng2015,SagiOreg2015,Macieiko2010} in the regime of strong electron-electron interactions. The main idea of the analysis is to search for solutions minimizing the energy of the system, which translates into maximizing the size of the gap opened by backscattering-assisted tunneling processes  and should stabilize the system, similar to nesting conditions discussed before in various systems~\cite{nesting_1,nesting_2,nesting_3,nesting_4}. Importantly, the condensation energy gain is maximum for processes that do not break the rotational and translation symmetries of the system \cite{Affleck1,Affleck2}. This helps us to reduce the problem effectively to one dimension where we can then use bosonization and Luttinger liquid techniques to show the existence of fractionally charged surface states with a non-degenerate helical Dirac cone spectrum in the topological phase. 
Additionally, in Section~\ref{WSM} we discuss how an equal combination of Rashba and Dresselhaus SOI leads to a Weyl semimetal phase in non-interacting systems. We summarize our results in Section VI.

		\begin{figure}[!h]
			\centering
			\includegraphics[width=0.75\columnwidth]{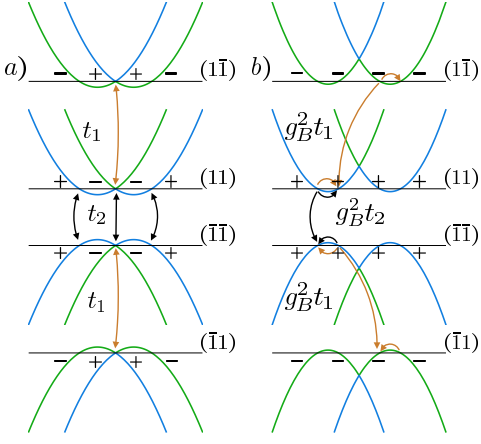}
			\caption{Dispersion relation of the layers for a fixed value of $\theta$. (a) The chemical potentials $\mu$ (black lines) are tuned to the SOI energy $E_{so}$. The colors blue/green encode positive/negative helicity. The arrows represent the tunneling processes between fields  allowed by spin and momentum conservation laws. (b) The chemical potential is tuned to $E_{so}/9$. In the presence of strong 
			 interactions, tunneling processes assisted by backscattering dominate resulting in the bulk gap. The orange and black arrows represent terms in $\mathcal{O}_{t_1}$ and $\mathcal{O}_{t_2}$, respectively [see Eqs. (\ref{BackscatteringAssistedT1}) and (\ref{BackscatteringAssistedT2})].}
			\label{Tunnelings}
		\end{figure}		
		
\section{Model}\label{Model}
We consider a system of weakly coupled two-dimensional electron gas (2DEG) layers stacked along the $z$ axis with the distance $a_z$ between layers. The unit cell consists of four layers, two of which have an electron-like dispersion and two have a hole-like  dispersion, see Fig.~\ref{setup}. Each layer has a SOI of Rashba type.  The strength of the SOI, $\alpha$, is the same throughout the unit cell but alternates its sign from layer to layer. 
We introduce two indices to label the layers: $\eta \in \{1,\bar{1}\}$ and $\tau \in \{1,\bar{1}\}$, which distinguish between electron and hole layers as well as between layers with positive and negative SOI, respectively.
The kinetic part of the Hamiltonian reads
\begin{align}
	&H_0 =  \sum^N_{n=1} \sum_{\eta \tau} \sum_{\sigma \sigma'}\int  dx dy \ \Psi^{\dagger}_{n \eta \tau \sigma} h^0_{ \eta \tau \sigma \sigma'} \Psi_{n \eta \tau \sigma'}  \label{UncoupledHamiltonian} \\	
	 &h^0_{ \eta \tau \sigma \sigma'} =  \eta \left( - \frac{\hbar^2}{2m} \nabla^2 - \mu \right) -i\tau \alpha (\sigma_1 \partial_y - \sigma_2 \partial_x)_{\sigma\sigma'}, \label{UncoupledHamiltonianDensity}
	\end{align}
where the sum runs over $N$ unit cells and  $\Psi_{n \eta \tau \sigma}(x,y)$ is the annihilation operator of an electron in the $(\eta \tau)$-layer  of the $n$-th unit cell with spin-projection $\sigma=\pm 1$ at position $(x,y)$. The chemical potential $\mu$ is measured from the SOI energy $E_{so}=m \alpha^2/2 \hbar^2$ in each layer and has the same magnitude in all layers. The dispersion of the $(\eta \tau)$-layer is given by $E_{\pm}(k)= \eta \frac{\hbar^2 k^2}{2m} \pm \tau \alpha k$ with the eigenstates characterized by spinor $\ket{\eta,\tau,\sigma, \theta} =  \left(1, - i \eta \tau \sigma  e^{i \theta} \right)^T/\sqrt{2}$, where $\theta$ is the angle between the 2D momentum vector $\mathbf{k}$ and the $k_x$ axis.

In the following, we consider spin-conserving tunneling  between layers.
For a tunneling process of amplitude $t_1$ ($t_2$) between layers of the same (opposite) mass the Hamiltonian is given by
	\begin{align}
&	{H}_{t_2} =  t_2 \sum_{n \sigma} \int dx dy \left( \Psi^{\dagger}_{n 1 \bar{1} \sigma} \Psi_{(n-1) \bar{1} 1 \sigma} + \Psi^{\dagger}_{n\bar{1}\bar{1}\sigma} \Psi_{n11\sigma} + \text{H.c.} \right), \nonumber \\
&{H}_{t_1} =  t_1 \sum_{n \sigma \eta } \int dx dy \left( \Psi^{\dagger}_{n \eta 1 \sigma} \Psi_{n \eta \bar{1} \sigma} + \text{H.c.}\right).
	\label{Ht2}
	\end{align}
Without loss of generality, we assume that $t_1,t_2\geq 0$.

\section{THREE DIMENSIONAL STRONG TOPOLOGICAL INSULATOR} \label{3DTI}

\subsection{Bulk Spectrum and Symmetry Class}
To begin with, we show that there is a bulk gap at $\mu=0$ and the symmetry class the Hamiltonian falls into is AII. In order to do so, we first consider an infinite system and
 introduce momenta ($\bf k$, $k_z$).
  The total Hamiltonian is given by $H=H_0+H_{t_1}+H_{t_2}$ [see Eqs. \eqref{UncoupledHamiltonian}-\eqref{Ht2} and \eqref{HamiltonianDensityPauliForm}] with the exact bulk spectrum 
	\begin{align}\label{BulkSpectrum}
	&E^2 _{\pm}(k,k_z)= \epsilon^2 + (\alpha k)^2 + (t_1^2+t_2^2) \nonumber \\
	&\hspace{50pt} \pm 2\sqrt{(\alpha k)^2 \epsilon^2 + t_1^2\epsilon^2+t_1^2t_2^2 \cos^2(k_za_z/2)},
	\end{align}
	where $\epsilon(k)= \hbar^2 k^2/2m$.  If $t_1\neq t_2$ and $t_2>0$, the bulk is fully gapped. If $t_1= t_2>0$, the bulk gap closes at $({\bf k}, k_z) =(0,0)$. The system can be tuned into topological ($t_1>t_2$) and trivial ($t_2>t_1$) phase, as shown below.
	
	In order to discuss the symmetry class of the Hamiltonian we rewrite the total Hamiltonian in terms of Pauli matrices  $\sigma_i$, $\eta_i$,  and $\tau_i$ acting in spin and layer space, respectively. As a result, we obtain
	\begin{widetext}
		\begin{align}
		&H= \int d{\bf k} dk_z \ \Psi^{\dagger}(\mathbf{k},k_z) h (\mathbf{k},k_z) \Psi(\mathbf{k},k_z), \label{HamiltonianInPauliForm}\\
		&h(\mathbf{k},k_z)= \Big[\frac{\hbar^2 (k_x^2+k_y^2)}{2m} -\mu \Big]\eta_3 + \alpha \tau_3 (\sigma_1 k_y - \sigma_2 k_x) +t_1 \tau_1 \nonumber \\
		&\hspace{100pt}+ t_2 \Big[\cos^2(k_z a_z /2) \eta_1 \tau_1 + \sin^2(k_z a_z /2) \eta_2 \tau_2
		+  \sin(k_z a_z )(\eta_1 \tau_2 + \eta_2 \tau_1)/2 \Big]. \label{HamiltonianDensityPauliForm}
		\end{align}
	\end{widetext}
	The Hamiltonian is invariant under time reversal operation 
	$\Theta = i \sigma_2 \mathcal{K}$, where $\mathcal{K}$ is the complex conjugation operator. In the Altland-Zirnbauer classification \cite{AltlandZirnbauer1997,Zirnbauer1996} there are three symmetry classes with $\Theta^2=-1$, two of which have additional particle-hole symmetry. Under particle-hole transformation $\mathcal{C}$ the Bloch Hamiltonian has to satisfy $\mathcal{C} h^T(-\mathbf{k},-k_z) \mathcal{C}^{-1} = - h(\mathbf{k},k_z)$. No such operator could be found for $h(\mathbf{k},k_z)$ [see Eq. \eqref{HamiltonianInPauliForm}] and it therefore belongs to the AII symmetry class \cite{AltlandZirnbauer1997,Zirnbauer1996}. In three dimensions, the system is classified by a $\mathbb{Z}_2$ invariant and can be a strong topological insulator hosting single helical Dirac cones at each surface \cite{RyuSchnyder2010}.
	
We end this subsection by computing the topological invariant explicitly following Ref. \cite{RyuSchnyder2010}. We derive an effective Hamiltonian by identifying the low-energy bands that close at the phase transition and the level crossing wave vector and then expand the Hamiltonian around these points.
From our analysis we know that the topological phase transition takes place at $t_1=t_2$ and the gap closes at $(\mathbf{k},k_z)=(0,0)$. Expanding Eq. \eqref{HamiltonianDensityPauliForm} around this momentum point and performing a unitary transformation, the Hamiltonian can be brought into a form consisting of two decoupled $4 \times4$ blocks containing the low-energy and high-energy bands. The low-energy bands undergo the topological phase transition at $t_1=t_2$. Projecting onto the subspace containing these low-energy bands that close at the critical point, one obtains the effective Hamiltonian given by
\begin{equation}
h_{\text{eff}}(\mathbf{k},k_z) = \alpha k_x  \sigma_2 - \alpha k_y  \sigma_1- \frac{a_z t_2}{2}k_z \eta_2  \sigma_3 + M \eta_1  \sigma_3 , \label{12}
\end{equation}
where we introduced the mass $M=t_1-t_2$. Using this simplified Hamiltonian, one can calculate the  $\mathbb{Z}_2$ invariant $\nu_0$ explicitly. We find that $\nu_0=1$ ($\nu_0=0$) if $M>0$ ($M<0$). We note that the same Hamiltonian was studied before in Ref.  \cite{Qi2008}, where it was shown that $h_{\text{eff}}(\mathbf{k},k_z) $ [Eq. (\ref{12})] corresponds to a 3D strong TI.
	
\subsection{Existence of Surface States}	
The presence of one helical Dirac cone at {\it any} surface is a central feature of a strong TI. In this subsection, we show the existence of these surface states in the topological regime $t_1>t_2$ using analytical and numerical methods. First, we prove the existence of surface states on the top and bottom surface of the stack, i.e. at the boundaries orthogonal to the stacking direction. To this end, we restrict the discussion to the low-energy regime and perform a linearization of the Hamiltonian by assuming $t_1, t_2 \ll E_{so}$. We represent the momentum in polar coordinates to exploit the rotational symmetry of the layers.
Being functions of good quantum numbers, the modulus $k$ and the polar angle $\theta$ are constants for $k_x$ and $k_y$ fixed. The spectrum of each layer is isotropic and, therefore, independent of $\theta$, which means that there are only fluctuations in $k$ direction. States with different $\theta$ are decoupled from each other and we can treat the problem as effectively one-dimensional in direction of propagation $r$~\cite{Luka2016}.

We linearize the spectrum and represent the field operators around the Fermi surface (FS) in terms of slowly varying fermionic field operators $S^{\delta,\beta}_{n \theta \eta \tau }(r)$ \cite{Luka2016,composite},
	\begin{equation}\label{linearizedField}
	\Psi_{n \theta \eta \tau \sigma}(x,y) = \sum_{\delta \beta} \alpha_{\delta \beta \theta \eta \tau \sigma} S^{\delta,\beta}_{n \theta \eta \tau } e^{i \eta \beta k^{\delta}_F(x \cos\theta+y \sin\theta)},
	\end{equation}
where $\delta= e,i$ labels  the  exterior ($e \equiv 1$) and interior ($i\equiv \bar1$) FS, with corresponding Fermi momenta  $k_F^e=2 m \alpha/ \hbar^2$ and $k_F^i=0$. Here, $\beta=1, \bar{1} $ refers to `right' ($1$) and `left' ($\bar 1$) movers propogating into opposite $r$-directions.  In this representation the polar angle $\theta$ is restricted to $[0,\pi)$. The spin overlap amplitude is given by
	$\alpha_{\delta \beta \theta \eta \tau \sigma}= \braket{\sigma| \eta,\tau,\delta,\theta+\pi(1-\eta\beta)/2}$.
The linearized kinetic term [see Eq. \eqref{UncoupledHamiltonian}] becomes $\bar{H}_0 = \int   dr\  \mathcal{\bar{H}}_0 (r) $ with
	\begin{equation}\label{LinearizedRashbaHamiltonian}
	\bar{\mathcal{H}}_0= -i \hbar v_F \sum_{n=1}^N \sum_{ \theta \eta \tau  } \sum_{\delta \beta} \beta (S^{\delta,\beta}_{n \theta \eta \tau})^{\dagger}(\partial_r S^{\delta,\beta}_{n \theta \eta \tau}),
	\end{equation}
where $v_F=\alpha/\hbar$ is the Fermi velocity.

\begin{figure}[t]
		\centering
		\includegraphics[width=0.9 \columnwidth]{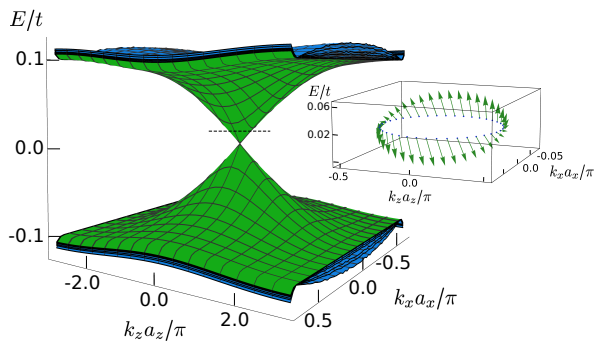}
		\caption{The spectrum in the topological phase obtained numerically for $N_y=300$, $t_1/t=0.2$, $t_2/t=0.1$, $\bar{\alpha}/t=0.3$, see SM~\cite{SM}. The bulk states (blue) are fully gapped with gap  $\Delta_{\text{min}}=2(t_1-t_2)$. The dispersion of the surface states localized in the $xz$ plane (green) is represented by an anisotropic Dirac cone. The inset shows the helical spin structure in the layer $\eta=\bar\tau=1$ at  $y/a_y=1$ for $E/t=0.02$ (dashed line), confirming the presence of a single {\it helical} Dirac cone at the $xz$-surface. }
		\label{DiracCone}
	\end{figure}

We next employ a two-step perturbation approach by considering the regime $t_1\gg t_2$ \cite{Klinovaja2015}. 
We first take into account the $t_1$ tunnelings and obtain after linearization,
	\begin{equation}\label{t1hopping}
	\bar{\mathcal{H}}_{t_1} = t_1 \sum_{n=1}^N \sum_{ \theta \eta \beta} \Big[ (S^{i,\beta}_{n \theta \eta 1})^{\dagger}S^{i,\bar{\beta}}_{n \theta \eta \bar{1}} +\text{H.c.} \Big].
	\end{equation}
Importantly, $\bar{\mathcal{H}}_{t_1}$ couples fields of opposite velocities at ${\bf k}=0$ 
resulting in a partial gap (see also Fig. \ref{Tunnelings}a). In a next step, we take into account the fields that are unaffected by the $t_1$-term and neglect the already gapped fields. Analogously, one obtains
	\begin{equation}\label{t2hopping}
	\hspace{-5pt}	\bar{\mathcal{H}}_{t_2} = t_2 \sum_{\theta  \beta }  \Big[ \sum_{n=2}^N (S^{e,\beta}_{n \theta  1\bar{1}} )^{\dagger} S^{e,\bar{\beta}}_{(n-1) \theta \bar{1}1} +  \sum_{n=1}^N (S^{e,\beta}_{n \theta \bar{1} \bar{1}})^{\dagger} S^{e,\bar{\beta}}_{n \theta 11} +\text{H.c.} \Big].
	\end{equation}
These terms gap out the remaining fields in the bulk  but do not affect the two fields at the top, $S^{e,1/\bar{1}}_{1 \theta 1 \bar{1}}$, and at the bottom, $S^{e,1/\bar{1}}_{N \theta \bar{1} 1}$, of the stack, since they do not appear in Eq. \eqref{t2hopping}. These surface states are gapless, have a linear dispersion and the spin  of each state is locked to be orthogonal to its momentum,
in other words, they form a single helical Dirac cone at  each of the two surfaces. We remark that starting the perturbative analysis in the opposite regime $t_2 \gg t_1$ all fields at the top and bottom surfaces are gapped and the system is in the trivial insulating state. The result obtained in the perturbative regimes smoothly connects to the region of the phase diagram where $t_1 > t_2$, and due to their topological nature the gapless surface states persist over the whole parameter range, see Fig. \ref{DiracCone}.

\begin{figure}[b]
	\centering
	\includegraphics[width=0.9\columnwidth]{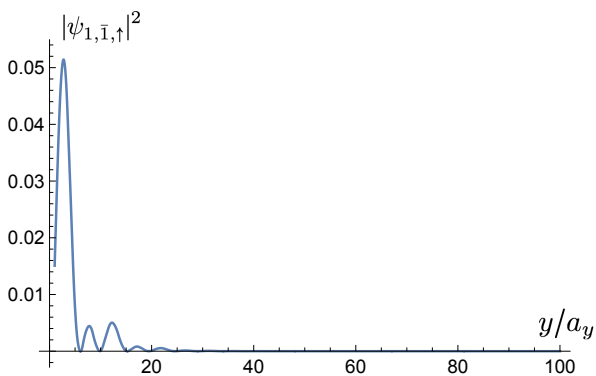}
	\caption{The probability density of the wavefunction $|\psi_{1, \bar{1}, \uparrow}|^2$ in the ($1, \bar{1}$) layer for a state on the Dirac cone on the first one hundred sites ($N_y=800$) along $y$ direction. The figure was obtained for $t_1/t=0.2$, $t_2/t=0.1$, $\bar{\alpha}/t=0.3$, and $\mu=-4t$ for a state on the Dirac cone with $k_x a_x/\pi = 0.05$ and $k_z a_z/\pi=0$. The wavefunction is localized at the $xz$ surface decaying rapidly along $y$ direction into the bulk.}
	\label{WaveFunction}
\end{figure}

To access the spectrum of the, say, $xz$ surface at $y=0$, we employ numerical diagonalization (see Appendix for more details) and consider the system finite in $y$-direction with $N_y$ lattice sites. The spectrum of the tight-binding Hamiltonian [see Eq. \eqref{TightBindingHamiltonian}] in the topological phase is shown in Fig.~\ref{DiracCone}. It can be seen that the bulk states are separated by an energy gap and that there exist states with a Dirac spectrum. In Fig.~\ref{WaveFunction}, we show the modulus squared of the wavefunction of a state on the Dirac cone on the first hundred lattice sites. One can observe that, indeed, the state is localized at the surface of the system and therefore conclude that the Dirac cone corresponds to surface states on $xz$ surface. The spin is locked to the momentum resulting in a helical texture (see the inset in  Fig.~\ref{DiracCone}). Since the system has rotational symmetry around the $z$ axis, it is clear that the same conclusions could be drawn if we had imposed a hard wall boundary condition at $x=0$. In conclusion, we showed the existence of a single helical Dirac cone on each boundary and a fully gapped bulk spectrum in the regime $t_1>t_2$. We emphasize that the gapless surface states were obtained in a non-perturbative regime which proves that their existence does not rely on the perturbative approach considered above, underlining their topological nature.

\begin{figure}[t]
	\centering
	\includegraphics[width=.8\columnwidth]{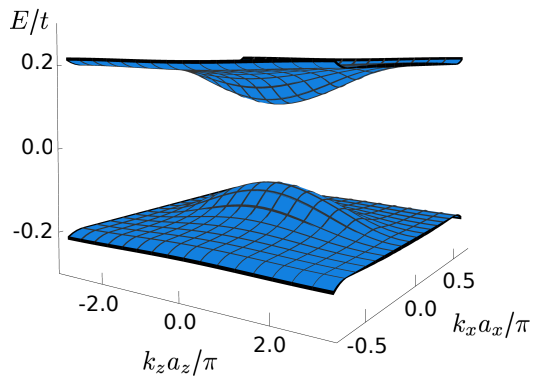}
	\caption{The spectrum in the trivial phase obtained numerically for $N_y=300$, $t_1/t=0.1$, $t_2/t=0.2$, $\bar{\alpha}/t=0.3$. The spectrum is fully gapped and no surface states were found in the $xz$ plane. The same holds for all other surfaces, which shows that here the system is in the trivial phase.}
	\label{SpectrumTrivial}
\end{figure}

For completeness we also show the spectrum in the region $t_1<t_2$ which is separated by the gap closing line $t_1=t_2$ from the region where we found the Dirac cone. As can be seen in Fig. \ref{SpectrumTrivial} the spectrum is fully gapped and  there are no surface states. The system is a trivial insulator in the whole region where $t_1<t_2$. This analysis confirms that for $t_1>t_2$ ($t_1<t_2$), the strong topological  $\mathbb{Z}_2$ invariant $\nu_0$ is given by $\nu_0=1$ ($\nu_0=0$), which is consistent with our analysis above.

\begin{figure}[b]
	\centering
	\includegraphics[width=.9\columnwidth]{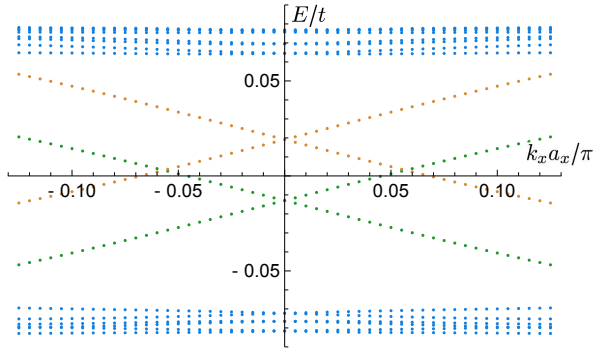}
	\caption{The spectrum of the topological phase in the presence of disorder along the cut $k_z a_z/\pi=0$ with parameters $N_y=300$, $t_1/t = 0.2$, $t_2/t=0.1$,  $\bar{\alpha}/t=0.3$, $\mu= -4t$, and $\sqrt{\braket{\delta \mu^2}} = 0.7 t_2$. The two Dirac cones on two opposite surfaces (green and orange) still exist and there is no gap opened by disorder. However, due to slightly different disorder configurations at two surfaces, there is a small shift of the position of the center of the Dirac cone in energy.}
	\label{DiracConeDisorder}
\end{figure}

To emphasize the topological origin of surface states, we show their stability against non-magnetic disorder. In order to do so, we modified our tight-binding model to implement detuning of the chemical potential in the ($\eta$,$\tau$)-layer at lattice site $i$ by $\delta \mu_{\eta \tau i}$. The perturbations were randomly generated from a normal distribution centered at $\braket{\delta \mu}=0$ and the variance characterizing the disorder strength was chosen such that $\sqrt{\braket{\delta \mu^2}}<t_2$. As can be seen from Fig.~\ref{DiracConeDisorder}, the surface states remain intact in the presence of non-magnetic disorder. There is no gap opening. Moreover, this analysis also confirms that our initial assumption of the rotational and translational symmetries is not crucial for the existence of the topological phase.

	\section{Fractional Topological Insulator}\label{FTI}
\begin{figure}
		\centering
		\includegraphics[width=.5\columnwidth]{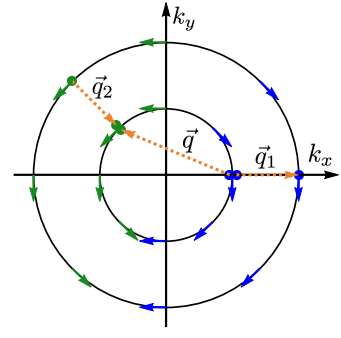}
		\caption{Schematics of backscattering assisted tunneling process between two neighboring electron layers. The two Fermi surfaces are drawn on top of each other.  For brevity, the spin polarizations of the corresponding state at the Fermi surface for the first (green arrows) and second layer (blue arrows) is indicated only for $k_x<0$ ($k_x>0$).
		 The involved electrons residing in the respective layers are shown by blue (green) dots.
The momentum transfer $\vec{q}$ (orange dotted arrows) during the tunneling event is compensated by the two backscattering processes such that $\vec{q}+\vec{q}_1+\vec{q}_2=0$. Correspondingly, due to the spin structure of the Fermi surface, the amplitude of the  tunneling process with momentum transfer $\vec{q}$ connecting two states with misaligned spins is reduced. The backscattering assisted tunneling amplitude (and thus the resulting bulk gap) is maximum if all involved spins are aligned, thus, $\vec q \parallel \vec{q}_{1,2}$. In this case, the rotation symmetry of the system is preserved. }
		\label{BackscatteringAssistedTunnelingProcess}
	\end{figure}

	Next, our goal is to identify the regime in which the system is a fractional strong 3D TI. For this, we detune the chemical potential in $H_0$ [see Eq. (2)] to $\mu_{1/3} = E_{so}/9$. This particular choice of $\mu$ fixes the ratio between the radii of the interior and exterior FS to $2k_F^i=k_F^e=4 m \alpha/3 \hbar^2$. We, again, restrict the discussion to the regime $t_1 \gg t_2$ and treat the $t_1$-terms first. The direct tunneling ($t_1$) between layers of the same mass is forbidden by spin/momentum conservation and do not result in a gap. Repulsive electron-electron interactions, however, open the channel for backscattering assisted tunneling which has a chance to open a gap. These processes consist of a tunneling with non-zero momentum transfer which is accompanied by two backscattering events (in leading order) ensuring overall momentum conservation. If the tunneling occurs between two states where the spins are misaligned, the tunneling amplitude gets suppressed by a factor of the spin overlap (see Fig. \ref{BackscatteringAssistedTunnelingProcess}). 
	Thus, we only take into account events where the tunneling amplitude and correspondingly the size of the bulk gap becomes maximal \cite{nesting_1,nesting_2,nesting_3,nesting_4}, which corresponds to processes preserving the rotational and translational symmetries of the system and do not mix states characterized by different values of $\theta$ \cite{,Affleck1,Affleck2}, see Fig. \ref{Tunnelings}b. Such processes similar to nesting conditions on Fermi surfaces~\cite{nesting_1,nesting_2,nesting_3,nesting_4} allow us to maximize the condensation energy gain (also known as Peierls-type energy gain) and stabilize the topological phase~\cite{nesting_1,nesting_2,nesting_3,nesting_4}. If the chemical potential is detuned by $\delta \mu$, the tunneling no longer conserves momentum exactly. However, the gap is still opened, although suppressed, if $\delta \mu <t_1,t_2$.
	
	The Hamiltonian density describing tunneling  between layers of the same mass becomes (see also Fig. \ref{Tunnelings}b)
	\begin{align}\label{BackscatteringAssistedT1}
	\hspace{-8pt} \mathcal{O}_{t_1} =g_1 \sum_{n \theta \eta \tau} \big[ &(S^{e,1}_{n\theta \eta \bar{\tau}})^{\dagger}S^{i,\bar{1}}_{n\theta \eta \bar{\tau}} (S^{i,1}_{n\theta \eta \tau})^{\dagger} S^{i,\bar{1}}_{n\theta \eta \bar{\tau}} (S^{i,1}_{n\theta \eta \tau})^{\dagger} S^{e,\bar{1}}_{n\theta \eta \tau} \nonumber \\
	& + \text{H.c.} \big],
	\end{align}
	with $g_1=t_1 g_B^2$ and $g_B$ being the strength of the backscattering term due to interactions. For the $t_2$ processes we distinguish the cases where tunneling occurs between the interior (exterior) FS. The operator that commutes with the one in Eq. \eqref{BackscatteringAssistedT1} is given in leading order by
	\begin{align}\label{BackscatteringAssistedT2}
	&\hspace{-5.8pt} \mathcal{O}_{t_2} = g_2  \sum_{\mathclap{\substack{n \theta \\ l \in \{1,\bar{1}\}}} } \big[ (S^{e,1}_{n \theta ll})^{\dagger} S^{i,\bar{1}}_{n \theta ll} (S^{e,1}_{n \theta ll})^{\dagger}S^{e,\bar{1}}_{n \theta \bar{l} \bar{l}} (S^{i,1}_{n \theta \bar{l} \bar{l}})^{\dagger} S^{e,\bar{1}}_{n \theta \bar{l} \bar{l}}   \\
	& \hspace{-6.5pt} + (S^{e,1}_{n_l \theta l \bar{l}})^{\dagger}S^{i,\bar{1}}_{n_l \theta l \bar{l}} (S^{e,1}_{n_l \theta \l \bar{l}})^{\dagger}S^{e,\bar{1}}_{\bar{n}_l \theta \bar{l} l} (S^{i,1}_{\bar{n}_l \theta \bar{l} l})^{\dagger} S^{e,\bar{1}}_{\bar{n}_l \theta \bar{l} l} +\text{H.c.} \big], \nonumber
	\end{align}
	with $g_2=t_2 g_B^2$ and where tunneling occurs between the exterior FSs. For brevity we use the index-dependent unit cell labels $n_l=n-(1-l)/2$ and $\bar{n}_l=n-(1+l)/2$.
	
For completeness we give the expression for the second operator describing $t_2$ tunneling processes between the interior FS of layers with opposite mass
\begin{align} \label{Bosonizedt2NonCom}
&\tilde{\mathcal{O}}_{t_2} = g_2  \sum_{\mathclap{\substack{n \theta \\ l \in \{1,\bar{1} \}}}} \big[(S^{e,1}_{n \theta l l})^{\dagger} S^{i,\bar{1}}_{n \theta l l}  (S^{i,1}_{n \theta \bar{l} \bar{l}})^{\dagger} S^{i,\bar{1}}_{n \theta ll}  (S^{i,1}_{n \theta \bar{l} \bar{l}})^{\dagger} S^{e,\bar{1}}_{n \theta \bar{l} \bar{l}} \nonumber \\
&+ (S^{e,1}_{n_l \theta l \bar{l}})^{\dagger} S^{i,\bar{1}}_{n_l \theta l \bar{l}}  (S^{i,1}_{\bar{n}_l \theta \bar{l} l})^{\dagger} S^{i,\bar{1}}_{n_l \theta l \bar{l}}  (S^{i,1}_{\bar{n}_l \theta \bar{l} l})^{\dagger} S^{e,\bar{1}}_{\bar{n}_l \theta \bar{l} l} + \text{H.c.}\big],
\end{align}
with $n_l$ and $\bar{n}_l$ as above. While $\mathcal{O}_{t_1}$ and $\mathcal{O}_{t_2}$ commute, $\tilde{\mathcal{O}}_{t_2}$ does not commute with $\mathcal{O}_{t_1}$ and therefore these operators can not be diagonalized simultaneously. Thus, they leave the system gapless and, consequentially, do not result in an energy gain. Such terms can therefore be dropped.

The terms $\mathcal{O}_{t_1}$ and $\mathcal{O}_{t_2}$ open a gap in the bulk spectrum but in order to access the nature of the gapless surface states we employ the bosonization procedure for 1D systems. This is justified since in the limit of dominant tunneling, fields with different angles $\theta$ are not coupled [cf. Eqs. \eqref{BackscatteringAssistedT1} and \eqref{BackscatteringAssistedT2}]. Thus, for $\theta$ fixed, the problem is equivalent to tunnel-coupled infinite wires.

	We next introduce chiral bosonic fields 
	$\phi^{(\theta)}_{n \beta \eta \tau \sigma}(r)$  ($S^{\delta,\beta}_{n \theta \eta \tau} \sim e^{i \phi^{(\theta)}_{n \beta \eta \tau \sigma}}$), where $\beta$ determines the chirality and $\sigma$ the spin projection \cite{giamarchi}. 
	The chiral fields satisfy the commutation relation  $\Big[\phi^{(\theta)}_{n \beta \eta \tau \sigma}(r),\phi^{(\theta)}_{n \beta \eta \tau \sigma}(r') \Big] = i \pi \beta \text{sgn}(r-r')$
	and all other commutators vanish. In a more convenient basis defined as $\eta_{n \beta \eta \tau \sigma}^{(\theta)} = 2\phi_{n \beta \eta \tau \sigma}^{(\theta)} - \phi^{(\theta)}_{n \bar{\beta} \eta \tau \sigma}$,
	the above interaction terms become
	\begin{align}
	&\mathcal{O}_{t_2} = g_2 \sum_{\mathclap{\substack{\theta \\ l \in \{1, \bar{1}\}}} }  \left[ \sum_{n=1}^N\cos(\eta^{(\theta)}_{n 1 l l l} - \eta^{(\theta)}_{n \bar{1} \bar{l} \bar{l} l})+ \sum_{n=2}^{N}\cos(\eta^{(\theta)}_{n_l 1 l \bar{l} \bar{l}} - \eta^{(\theta)}_{\bar{n}_l \bar{1} \bar{l} l \bar{l}}) \right],\nonumber \\
	&\mathcal{O}_{t_1} = g_1 \sum_{n \theta \eta \tau}  \cos(\eta^{(\theta)}_{n \bar{1} \eta \bar{\tau} \bar{\tau}} - \eta^{(\theta)}_{n 1 \eta \tau \bar{\tau}}). \label{bosonizedt1} 
	\end{align}
	We work in the limit where $g_i$ are large compared to the quadratic part of the Hamiltonian 
	and the fields get pinned to one of the minima of the cosines \cite{giamarchi}. Using Eqs. \eqref{bosonizedt1}, we infer that all the fields in the bulk get pinned pairwise and, therefore, the bulk spectrum is fully gapped. However, as in the previous section, two fields at the top surface and two fields at the bottom surface do not appear in the tunneling terms, namely $\eta^{(\theta)}_{1,\bar{1} 1 \bar{1} \uparrow}$, $\eta^{(\theta)}_{1, 1 1 \bar{1} \downarrow}$, $\eta^{(\theta)}_{N,  1 \bar{1} 1 \uparrow}$, and $\eta^{(\theta)}_{N, \bar{1} \bar{1} 1 \downarrow}$. These fields do not get pinned and stay gapless \cite{Teo2014}. We conclude that the interacting system hosts gapless surface states that have their spin locked orthogonal to their momentum. The quasi-particle excitations on the surface are directly given by the exponential of the gapless bosonic fields listed above. These excitations have been shown to carry fractional charge $q=e/3$ \cite{Oreg2014}. This procedure can be generalized to other odd integers $n>3$ to obtain fractional TIs with $q=e/n$.
	
		\begin{figure}[t]
		\centering
		\includegraphics[width=\columnwidth]{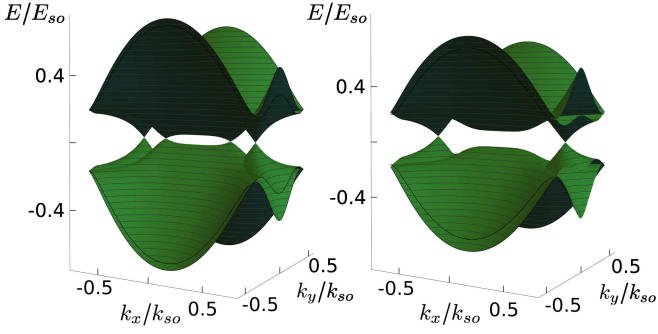}
		\caption{The Weyl semimetal spectrum 
			for the parameters $t_1/E_{so}=0.3$, $t_2/E_{so}=0.19$, and $k_z=0$. \emph{Left:} Four Weyl nodes exist for weak exchange interaction ($J/E_{so}=0.08$). \emph{Right:} In the regime $|J-t_1|<t_2$, only two Weyl outer nodes remain ($J/E_{so}=0.17$).}
		\label{WeylNodes}
	\end{figure}

\section{Weyl Semimetal phase}\label{WSM}
Remarkably, if the SOI is no longer of pure Rashba type but given by an equal combination of Rashba and Dresselhaus SOI by gate tuning \cite{Nitta1997}, such that the spin gets coupled only to the momentum in one particular direction \cite{Leo2017,Schliemann2003,Bernevig2006,Meier2007,Koralek2009,
Walser2012,Dettwiler2014,MengKlinovaja2014}, a Weyl semimetal phase (WSM) can be realized in the same setup. The modified SOI term reads
\begin{equation} \label{WeylSOI}
\hspace{-8pt}	\tilde{H}_{\text{SOI}}= -i \alpha \sum_{n \eta \tau} \sum_{\sigma  \sigma'} \int dx dy \ \Psi^{\dagger}_{n \eta \tau \sigma} \tau (\sigma_1)_{\sigma\sigma'} \partial_y \Psi_{n \eta \tau \sigma'}.
\end{equation}
As a result, the Hamiltonian density for each $\eta \tau$ layer is  given by
\begin{align}
\mathcal{\tilde{H}}_0(x,y) = \sum_{\substack{n \eta \tau \\ \sigma \sigma'}} &\Psi^{\dagger}_{n \eta \tau \sigma} \Big[\eta \left(- \frac{\hbar^2}{2m} \nabla^2 - \mu \right) -i\tau \alpha \sigma_1 \partial_y\Big]_{\sigma \sigma'} \Psi_{n \eta \tau \sigma'}.
\end{align}
In Refs.~\cite{Schliemann2003,Bernevig2006,Meier2007,Koralek2009,Walser2012,MengKlinovaja2014,Dettwiler2014} it was shown that combining Rashba and Dresselhaus SOI in a 2DEG can lead to a partial compensation of the two. The above term arises when Rashba and Dresselhaus SOI are of equal strength. The tunneling part of the total Hamiltonian $H_{t_1}+H_{t_2}$ remains the same. From the above equation it is evident, that spin projection along the $x$-axis is a good quantum number $\sigma=\pm 1$. The total $8 \times 8$ Bloch Hamiltonian is block diagonal and can be written in terms of Pauli matrices as
\begin{align}\label{WeylHamiltonian}
&\tilde{h}(\mathbf{k}) = \frac{\hbar^2(k_x^2+k_y^2)}{2m}  \eta_3    - \alpha  k_y \sigma_1      \tau_3  + t_1     \tau_1  \nonumber \\
&\hspace{20pt}+ t_2 \Big[ \cos^2( k_z a_z/2) \eta_1   \tau_1  +  \sin^2(k_za_z /2)  \eta_2   \tau_2 \nonumber \\
&\hspace{50pt} +  \sin(k_z a_z ) (\eta_1   \tau_2   +     \eta_2   \tau_1 )/2\Big].
\end{align}
From now on ${\bf k}$ denotes the 3D momentum and ${\bf k_{||}}$ the in-plane momentum. The energy spectrum of this Hamiltonian is found to be
\begin{align}\label{SpectrumWeyl}
\tilde{E}_{\pm}^2 &=\epsilon_{||}^2 + t_1^2+t_2^2 + (\alpha k_y )^2 \nonumber \\
& \hspace{8pt} \pm 2\sqrt{\epsilon_{||}^2 [t_1^2+(\alpha k_y )^2] + t_1^2t_2^2 \cos^2(k_z a_z /2)},
\end{align}
with $\epsilon_{||}= \hbar^2 k_{||}^2/2m$. If $t_1<t_2$,  the spectrum has a bulk gap.  For $t_1>t_2$ there are two doubly degenerate gapless bulk states at $\pm \mathbf{k}_D$ with $\mathbf{k}_D=(k^*_{x},0,0)$ and $k^*_{x} =  \sqrt{\frac{2m}{\hbar^2}}(t_1^2-t_2^2)^{1/4}$.  These are Dirac nodes hosting two Weyl nodes of opposite chirality at the same point. The two Weyl nodes are not coupled in the absence of disorder and, therefore, do not annihilate each other. Such Dirac nodes can, however, be stable only if additional crystal symmetries are present that stabilize these nodes \cite{DiracSM3D}.

\begin{figure}[b]
	\centering
	\includegraphics[width=.445\textwidth]{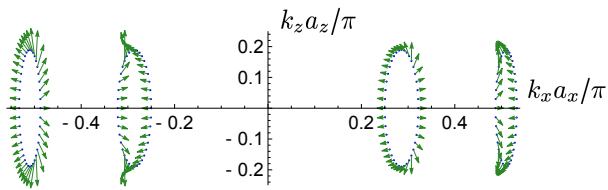}
	\caption{Spin structure of the states on the Weyl cones at an energy of $E/t=0.02$ at $y/a_y=260$ ($N_y=400$). For clarity we projected the spin onto the $k_x k_z$ plane. The parameters used for the numerics are $t_1/t=0.2$, $t_2/t=0.1$, $\bar{\alpha}/t=0.4$, and $J/t=0.07$.}
	\label{SpinStructureWeylNodes}
\end{figure}

Next, we would like to eliminate the twofold degeneracy of the Dirac node by splitting it into two Weyl nodes. This can be achieved if the time-reversal symmetry is broken, for example, via magnetic impurities which order  ferromagnetically (FM) along a direction orthogonal to the SOI direction, let say, in the $y$ direction~\cite{Burkov2011,Chen2010,Chang2015,Hosseini2015}. The exchange interaction between the electron spins and the magnetic impurities reads,
\begin{equation}
H_J = -J \sum_{n \eta \tau}\sum_{\sigma \sigma'}\int d^3 { x} \  \Psi^{\dagger}_{n \eta \tau \sigma} (\sigma_2)_{\sigma \sigma'} \Psi_{n \eta \tau \sigma'},
\end{equation}
with $J>0$. Gapless states only exist in the $k_z=0$ plane where the spectrum is given by
\begin{align}
\tilde{E}_{\pm,\pm,}(k_x,k_y,0)&=\epsilon_{||}^2 + (J \pm t_1)^2 + t_2^2 + (\alpha k_y)^2 \nonumber \\& \hspace{.5pt} \pm 2 \sqrt{(\epsilon_{||}^2+t_2^2)( J \pm t_1)^2 +\epsilon_{||}^2 (\alpha k_y)^2} .
\end{align}
If $|J-t_1|>t_2$ the bulk spectrum has four gapless states at $\pm \mathbf{k}_{\pm}$ with $\mathbf{k}_{\pm}=(k^*_{x,\pm},0,0)$ and $k^*_{x,\pm}=\sqrt{\frac{2m}{\hbar^2}}\Big[(J \pm t_1)^2-t_2^2 \Big]^{1/4}$. This means that the initial twofold degeneracy gets lifted and we end up with four distinct nodes (see Fig. \ref{WeylNodes}). The nodes at $\mathbf{k}_{\pm}$ and $-\mathbf{k}_{\pm}$ have opposite chirality (see Fig. \ref{SpinStructureWeylNodes}). If the exchange interaction strength is tuned to $J = t_1-t_2$, the two inner nodes meet and annihilate. As long as $|J-t_1|<t_2$, only the nodes at $\pm \mathbf{k}_+$ exist, while at $J=t_1+t_2$ the two inner nodes reappear and separate when $J$ is increased further. 

\begin{figure}
	\centering
	\includegraphics[width=0.8\columnwidth]{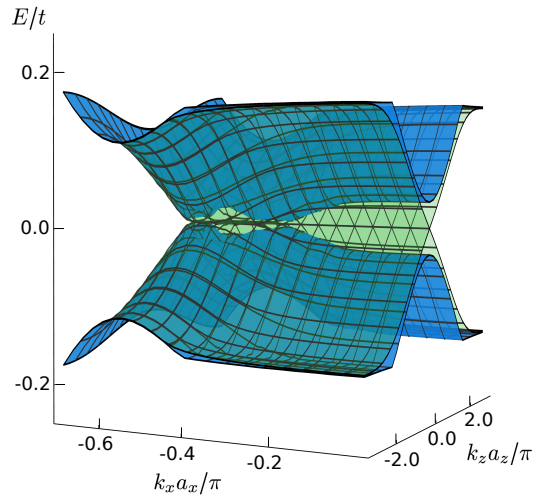}
	\caption{Spectrum in the Weyl semimetal phase in half of the BZ as a function of momenta ($k_x$, $k_z$). 
		The spectrum of semi-infinite system ($y>0$) was obtained numerically [cf. Eq. \eqref{WeylTightBinding}] for $N_y=800$, $\bar{\alpha}/t=0.45$, $t_1/t=0.3$, $t_2/t=0.19$, and $J/t=0.08$, i.e. in the regime $J<t_1-t_2$ where we expect four Weyl nodes from the analytical anlaysis. The bulk (surface) states are colored in blue (green). The spectrum is cut in half along the line connecting the gap closing points. The spectrum indeed features two (since we show only one half) gapless points as well as the Fermi arcs. The gap is not closed exactly due to finite size effects.}
	\label{SpectrumWeylSMTightBinding}
\end{figure}

In the Appendix B, we find explicitly spectrum and wavefunctions of surface states in the WSM phase. Here, we present the numerical spectrum obtained from a tight-binding model defined in Appendix A with polarized magnetic impurities in the regime $J<t_1-t_2$, see Fig.~\ref{SpectrumWeylSMTightBinding}. We confirm the existence of gapless bulk states and of surface states that are dispersionless in $x$ direction. The surface states have a linear dispersion in the $z$ direction as expected.
 In order to check the chirality of the Weyl nodes, we also access the spin structure of the states on the Weyl cones, see Fig. ~\ref{SpinStructureWeylNodes}.
Indeed, the overall chirality of four Weyl cones is zero. We note that the Weyl phase is defined strictly at the absence of disorder. Any finite disorder will scatter between Weyl cones as discussed in literature \cite{Burkov2011}.

\section{Conclusions} We considered a layered system that realizes a  3D fractional strong  TI. We constructed a simple model that solely consists of weakly coupled 2D layers with Rashba SOI. We also show that if Dresselhaus and Rashba  SOI term are of the same strength, the system can be brought into the Weyl semimetal phase. The motivation for such setups is given by the vast progress in fabricating superlattices and van der Waals heterostructures. 
We believe that these engineered materials provide a promising route towards realizing 3D (fractional) TIs and Weyl semimetals as proposed in this work.

\section*{ACKNOWLEDGEMENTS}
We acknowledge support  from the Swiss National Science Foundation and NCCR QSIT, and  the Marie Sklodowska-Curie Innovative Training Network (ITN-ETN) Spin-NANO.

\section*{APPENDIX A: TIGHT-BINDING MODELS}
In Sect.~\ref{3DTI}, we address the question of whether a helical Dirac cone exists at any boundary numerically by implementing a tight-binding model \cite{Luka2016} for a system with $N_y$ lattice sites. We impose a hard wall boundary at $y=0$ and consider the system to be infinitely extended along the $x$ and $z$ directions such that we can introduce momenta $k_x$ and $k_z$. The tight-binding Hamiltonian $H=\sum_{k_x k_z} H_{k_xk_z}$ for this setup reads
\begin{widetext}
	\begin{align}
	&H_{k_xk_z}= \sum_{\substack{ \eta \tau}} H_{0  k_xk_z \eta \tau } + \sum_{\substack{ i=1,2}}H_{t_i k_xk_z } , \label{TightBindingHamiltonian} \\
	&H_{0 k_x k_z \eta \tau} = - \sum_{n \sigma} \Big [  \eta \left(t \cos (k_x a_x) - \mu/2 \right) c^{\dagger}_{\eta \tau n \sigma} c_{\eta \tau n \sigma} +\eta t c^{\dagger}_{\eta \tau (n+1) \sigma} c_{\eta \tau n \sigma} \Big] \nonumber \\
	& \hspace{15pt}	+ \tau \tilde{\alpha} \sum_{n} \Big[i(c^{\dagger}_{\eta \tau (n+1) \uparrow} c_{\eta \tau n \downarrow} - c^{\dagger}_{\eta \tau (n-1) \uparrow} c_{\eta \tau n \downarrow})   + 2i \sin (k_x a_x) c^{\dagger}_{\eta \tau n \uparrow} c_{\eta \tau n \downarrow} \Big] +\text{H.c}.,\\
	&H_{t_1 k_x k_z} = t_1 \sum_{n \sigma} \Big [c^{\dagger}_{1 \bar{1} n \sigma} c_{11 n \sigma} + c^{\dagger}_{\bar{1} \bar{1} n \sigma} c_{\bar{1} 1 n \sigma} + \text{H.c.} \Big],\\
	&	H_{t_2 k_x k_z} = t_2 \sum_{n \sigma} \Big [ e^{-i k_z a_z}c^{\dagger}_{\bar{1} 1 n \sigma} c_{1 \bar{1} n \sigma} +  c^{\dagger}_{11 n \sigma} c_{\bar{1} \bar{1} n \sigma}  +\text{H.c.} \Big].
	\end{align}
\end{widetext}
Here, $c_{\eta \tau n \sigma}\equiv c_{\eta \tau n \sigma k_x k_y}$ is the annihilation operator for an electron with spin $\sigma$  in the layer $(\eta \tau)$ with momentum $(k_x, k_z)$ at position $y=n a_y$, where $a_y$ is the lattice constant along the $y$ direction. The spin-flip hopping amplitude is related to the SOI parameter by $\tilde{\alpha}= \alpha/2a_y$ (we take the lattice constants $a_x=a_y$) \cite{diego}. 

To describe the Weyl phase in Sect.~\ref{WSM}, we use the same Hamiltonian as in Eq. \eqref{TightBindingHamiltonian} besides  adding exchange interaction term and modifying the SOI term, such as
\begin{widetext}
	\begin{align}\label{WeylTightBinding}
	\tilde{H}_{0 k_x k_z \eta \tau} &= - \sum_{n \sigma} \Big [  \eta \left(t \cos (k_x a_x) - \mu/2 \right) c^{\dagger}_{\eta \tau n \sigma} c_{\eta \tau n \sigma} +\eta t c^{\dagger}_{\eta \tau (n+1) \sigma} c_{\eta \tau n \sigma} \Big] \nonumber \\
	&\hspace{90pt}	+i \tau \tilde{\alpha} \sum_{n} \Big[c^{\dagger}_{\eta \tau (n+1) \uparrow} c_{\eta \tau n \downarrow}  - c^{\dagger}_{\eta \tau (n-1) \uparrow} c_{\eta \tau n \downarrow}  \Big]  - i J \sum_n c^{\dagger}_{\eta \tau n \uparrow} c_{\eta \tau n \downarrow} +\text{H.c.},
	\end{align}
\end{widetext}
with the magnetic impurities of the strength $J$ polarized along $y$ direction. 

\section*{APPENDIX B: Analytical Calculation  of Surface States in Weyl Semimetal phase}

In this Appendix we explicitly show the analytical calculation of the surface states that appear in the WSM phase. We restrict the discussion to the regime where $t_1-t_2<J<t_1+t_2$ with $t_1>t_2$. Without loss of generality all three parameters are considered to be positive. In this range of the exchange interaction strength there exist two Weyl nodes at $\pm \mathbf{k}_+$ (see above) and the associated Fermi arcs are located on the $xz$ and $xy$ surface BZ. For simplicity we calculate the surface states on the $xz$ surface for $y=0$.  
In order to perform the  linearization  of the Hamiltonian [see Eq. \eqref{WeylHamiltonian}] we again assume $t_1, t_2\ll E_{so}$ (see also text above). The chemical potential is tuned to the Weyl nodes.

We start with the full 3D bulk Hamiltonian $\tilde{h}(k_x,k_y,k_z)$. Since the Fermi arc is the line connecting the projections of the Weyl nodes on the corresponding surface BZ, it is clear that the surface states occur with $(k_x,0)$ in the $k_x  k_z$ plane. Before linearization the Hamiltonian reads
\begin{equation}\label{WeylHamiltonianBeforeLinearization}
\tilde{h}(k_x,k_y,0) = \frac{\hbar^2 }{2m}(k_x^2+k_y^2) \eta_3 - \alpha k_y \tau_3 \sigma_3 + t_1 \tau_1 + t_2 \eta_1 \tau_1 + J \sigma_2.
\end{equation}
For a clean surface $(k_x,k_z)$ are good quantum numbers and we can solve Eq. \eqref{WeylHamiltonianBeforeLinearization} for fixed $k_x$, $k_z$, i.e. the problem reduces to a one-dimensional problem solving for the zero-energy eigenstates of $\tilde{h}_{k_x}(k_y\to-i \partial_y)$ ($k_z$ is fixed to zero). Since $k_x$ is fixed in Eq. \eqref{WeylHamiltonianBeforeLinearization} we treat the term containing $k_x$ as a detuning of the chemical potential from the spin-orbit energy. This procedure is justified since $k_x^2 \in [0, \frac{2m}{\hbar^2}\sqrt{(J+t_1)^2-t_2^2}]$ and therefore $k_x^2$ is bounded from above by $k_x^2< \frac{4\sqrt{2}m}{\hbar^2} t_1$. This yields $\delta \mu_{k_x} =\frac{\hbar^2 k_x^2}{2m}<  4\sqrt{2} t_1 \ll E_{so}$ in the perturbative regime $t_1$, $t_2 \ll E_{so}$.\\
Linearizing the Hamiltonian in Eq. \eqref{WeylHamiltonianBeforeLinearization} gives
\begin{align}\label{LinearizedWeylHamiltonian}
\tilde{h}&=  v_F\hat{k} \gamma_3  + t_1(\tau_1 \gamma_1- \tau_2 \gamma_2)/2 + t_2 \eta_1 \tau_1 \gamma_1 \nonumber \\ 
&\hspace{40pt}+ J(\sigma_2 \gamma_1 + \sigma_1 \tau_3 \gamma_2)/2 + \delta\mu_{k_x} \eta_3,
\end{align}
with $\hat{k}= -i \hbar \partial_y$ the momentum operator around the Fermi points and $\gamma_i$ acts in left/right mover space. The effect of spin-orbit coupling enters in two ways, firstly it determines the Fermi velocity (since $v_F=\alpha/\hbar$) and secondly it prevents the exchange interaction term from being diagonal in spin space, which would otherwise just lead to an energy shift of the two spin subbands and not produce any interesting effects. For the eigenstate we make the Ansatz $\psi_{k_x}(x,y,z)=e^{i k_x x} \psi(y)= e^{i k_x x} e^{i \lambda y} \phi_{\lambda}$, where $\phi_{\lambda}$ is a 16-component vector. Acting with the Hamiltonian in Eq. \eqref{LinearizedWeylHamiltonian} on $\psi_{k_x}$ one ends up with the matrix equation
\begin{equation}
\tilde{h}(k_x,\partial_y \to i \lambda) \phi_{\lambda} = E_{\lambda} \phi_{\lambda} .
\end{equation}
The surface states are the zero-energy eigenstates which decay for $y>0$ (this translates into the criterion $\text{Re}(i \lambda)<0$). The zero-energy states can simply be found by solving $\text{Det}(\tilde{h}(k_x,i\lambda))=0$. In the regime $t_1-t_2<J<t_1+t_2$ we find the following decaying solutions characterized by
\begin{align}
\lambda_{1,\pm} &= \frac{i t_2 \pm \delta\mu_{k_x}}{\hbar v_F}, \\
\lambda_{2,\rho,\kappa} &= \rho^{1-\kappa} i \frac{\sqrt{(J+\rho t_1)^2-\delta\mu_{k_x}^2} +\kappa t_2}{\hbar v_F},
\end{align}
where $\rho, \kappa \in \{-1,1\}$ and the corresponding eigenvectors (suppressing the normalization factors) are given by
\begin{align}
\phi_{1} &=(0, 0, 0, 0, 0, 0, 0, 0, i, 0, 0, 0, 0, 0, 0, 1), \nonumber \\
\phi_{2} &= (0, 0, i, 0, 0, 1, 0, 0, 0, 0, 0, 0, 0, 0, 0, 0),\nonumber  \\
\phi_{3} &= (0, 0, 0, 0, 0, 0, 0, 0, 0, 0, 0, -i, 1, 0, 0, 0), \nonumber \\
\phi_{4} &= (0, -i, 0, 0, 0, 0, 1, 0, 0, 0, 0, 0, 0, 0, 0, 0), \\
\phi_5 &= (i g_-^*,0,0,i,-1,0,0,g_-^*,0,-i,-i g_-^*,0,0,-g_-^*,1,0 ), \nonumber\\
\phi_6 &= (ig_-,0,0,i,-1,0,0,g_-,0,-i,-ig_-,0,0,-g_-,1,0), \nonumber \\
\phi_7 &= (-i g_+,0,0,i,1,0,0,g_+,0,i,-ig_+,0,0,g_+,1,0),\nonumber  \\
\phi_8 &= (ig_+,0,0,-i,1,0,0,g_+,0,-i,ig_+,0,0,g_+,1,0), \nonumber
\end{align}
with $g_{\pm} = \frac{\delta \mu_{k_x} - i \sqrt{(J \pm t_1)^2- \delta \mu_{k_x}^2}}{J \pm t_1}$. Some of these solutions seem to be ill-defined at $J=t_1$, but they actually have a finite limit once they are normalized. However, the expressions are too lengthy to be displayed here.\\
We write the general solution as linear combination in the basis $(\Psi_{1 \bar{1} \uparrow},\Psi_{1 1 \uparrow},\Psi_{\bar{1} \bar{1} \uparrow},\Psi_{\bar{1} 1 \uparrow},\Psi_{1 \bar{1} \downarrow},\Psi_{1 1 \downarrow},\Psi_{\bar{1} \bar{1} \downarrow},\Psi_{\bar{1} 1 \downarrow})$ and impose a hard-wall boundary condition at $y=0$. Dropping rapidly oscillating terms, we write
\begin{equation}
\psi(y) = \frac{1}{\sqrt{N}}\sum_{j=1}^8 c_j e^{i \lambda_j y} \tilde{\phi}_{\lambda_j},
\end{equation}
with $N$ as a normalization constant and $\tilde{\phi}$ an 8 component vector. We find a non-trivial zero energy solution characterized by coefficients ($c_5 =c_6=c_8=0 $),
\begin{equation}
c_2/c_1 = -c_7/c_1=1, \  c_3/c_1 =c_4/c_1= g_+.
\end{equation}

\bibliographystyle{unsrt}

\begin{thebibliography}{}
	\bibitem{KaneMele20051}
	C. L. Kane and E. J. Mele, Phys. Rev. Lett. \textbf{95}, 226801 (2005).
	
	\bibitem{KaneMele20052}
	C. L. Kane and E. J. Mele, Phys. Rev. Lett. \textbf{95}, 146802 (2005).
	
	\bibitem{Bernevig20061}
	C. Wu, B. A. Bernevig, and S. Zhang, Phys. Rev. Lett. \textbf{96}, 106401 (2006).
	
	\bibitem{Bernevig20062}
	B. A. Bernevig, T. L. Hughes, and S. Zhang, Science \textbf{314}, 5806 (2006).
	
	\bibitem{Koenig2007}
	M. K\"onig, S. Wiedmann, C. Br\"une, A. Roth, H. Buhmann, L. W. Molenkamp, X. Qi, and S. Zhang, Science \textbf{318}, 5851 (2007).
	
	\bibitem{FuKane2007}
	L. Fu, C. L. Kane, and E. J. Mele, Phys. Rev. Lett. \textbf{98}, 106803 (2007).
	
	\bibitem{Moore2007}
	J. E. Moore and L. Balents, Phys. Rev. B \textbf{75}, 121306(R) (2007).
	
	\bibitem{Hsieh2008}
	D. Hsieh, D. Qian, L. Wray, Y. Xia, Y. S. Hor, R. J. Cava, and M. Z. Hasan, Nature \textbf{452}, 970 (2008).
	
	\bibitem{Hsieh2009}
	D. Hsieh, Y. Xia, L. Wray, D. Qian, A. Pal, J. H. Dil, J. Osterwalder, F. Meier, G. Bihlmayer, C. L. Kane, Y. S. Hor, R.
	J. Cava, and M. Z. Hasan, Science \textbf{323}, 919 (2009).
	
	\bibitem{Levin2009}
	M. Levin and A. Stern, Phys. Rev. Lett. \textbf{103}, 196803 (2009).
	
	\bibitem{Maciejko2011}
	J. Maciejko, X. Qi, A. Karch, and S. Zhang, Phys. Rev. B \textbf{86}, 235128 (2011).
	
	\bibitem{KlinovajaTserkovniak2014}
	J. Klinovaja and Y. Tserkovniak, Phys. Rev. B \textbf{90}, 115426 (2014).
	
	\bibitem{Freedman20021}
	M.H. Freedman, M. J. Larsen, and Z. Wang, Commun. Math. Phys. \textbf{227}, 605 (2002).
	
	\bibitem{Cheng2012}
	M. Cheng, Phys. Rev. B \textbf{86}, 195126 (2012).
	
	\bibitem{Lindner2012}
	N. H. Lindner, E. Berg, G. Refael, and A. Stern, Phys. Rev. X \textbf{2}, 041002 (2012).
	
	\bibitem{Vaezi2013}
	A. Vaezi, Phys. Rev. B 87, 035132 (2013).
	
	\bibitem{Clarke2013}
	D. Clarke, J. Alicea, and K. Shtengel, Nat. Commun. 4,
	1348 (2013).
	
	\bibitem{KlinovajaLoss2014a}
	J. Klinovaja and D. Loss, Phys. Rev. Lett. \textbf{112}, 246403 (2014).
	
	\bibitem{KlinovajaLoss2014b}
	J. Klinovaja and D. Loss, Phys. Rev. B \textbf{90}, 045118 (2014).
	
	\bibitem{Kane2002}
	C. L. Kane, R. Mukhopadhyay, and T. C. Lubensky, Phys. Rev. Lett. \textbf{88}, 036401 (2002).
	
	\bibitem{Poilblanc1987} 
	D. Poilblanc, G. Montambaux, M. H\'eritier, and P. Lederer, Phys. Rev. Lett. \textbf{58}, 270 (1987).	
	
	\bibitem{Gorkov1995}
	L. P. Gor’kov and A. G. Lebed, Phys. Rev. B \textbf{51}, 3285 (1995).
	
	\bibitem{Klinovaja2013}
	J. Klinovaja and D. Loss, Phys. Rev. Lett. \textbf{111}, 196401 (2013).
	
	\bibitem{Teo2014}
	J. C. Y. Teo and C. L. Kane, Phys. Rev. B \textbf{89}, 085101 (2014).
	
	\bibitem{KlinovajaLoss2014c}
	J. Klinovaja and D. Loss, Eur. Phys. J. B {\bf 87}, 171 (2014).
	
	\bibitem{Meng2014}
	T. Meng, P. Stano, J. Klinovaja, and D. Loss, Eur. Phys. J. B \textbf{87}, 203 (2014).
	
	
	\bibitem{Neupert2014}
	T. Neupert, C. Chamon, C. Mudry, and R. Thomale, Phys. Rev. B \textbf{90}, 205101 (2014).
	
	\bibitem{Sagi2014}
	E. Sagi and Y. Oreg, Phys. Rev. B {\bf 90}, 201102 (2014).
	
	\bibitem{Klinovaja2015}
	J. Klinovaja, Y. Tserkovnyak, and D. Loss, Phys. Rev. B \textbf{91}, 085426 (2015).
	
	\bibitem{Peter} J. Klinovaja, P. Stano, and D. Loss, Phys. Rev. Lett. {\bf 116}, 176401 (2016).
	
	\bibitem{Sahoo2015}
	S. Sahoo, Z. Zhang, and J. C. Y. Teo, arXiv:1509.07133.
	
	\bibitem{Meng2015}
	T. Meng, Phys. Rev. B \textbf{92}, 115152 (2015).
	
	
	\bibitem{SagiOreg2015}
	E. Sagi and Y. Oreg, Phys. Rev. B \textbf{92}, 195137 (2015).
	
	\bibitem{Luka2016} 
	L. Trifunovic, D. Loss, and J. Klinovaja, Phys. Rev. B \textbf{93}, 205406 (2016).
	
	\bibitem{Nitta1997}
	J. Nitta, T. Akazaki, H. Takayanagi, and T. Enoki, Phys. Rev. Lett. \textbf{78}, 1335 (1997).
	
	\bibitem{Dettwiler2017}
	F. Dettwiler, J. Fu, S. Mack, P. J. Weigele, J. C. Egues, D. D. Awschalom, D. M. Zumb\"uhl, arxiv: 1702.05190 (2017).
	
	\bibitem{Novoselov2016}
	K. S. Novoselov, A. Mishchenko, A. Carvalho, and A. H. Castro Neto, Science \textbf{353}, 6298 (2016).
	
	\bibitem{Novoselov2013}	B. Sachs, L. Britnell, T. O. Wehling, A. Eckmann, R. Jalil, B. D. Belle, A. I. Lichtenstein, M. I. Katsnelson, and K. S. Novoselov, App. Phys. Lett. {\bf 103} 251607 (2013).
	
	\bibitem{Novoselov2012}	K. Novoselov, V. Fal'ko, L. Colombo, P. Gellert, M. Schwab, and K. Kim, Nature {\bf 490}, 192 (2012).
	
	\bibitem{smet}
	R. A. Deutschmann, W. Wegscheider, M. Rother, M. Bichler, G. Abstreiter, C. Albrecht, and J. H. Smet, Phys. Rev. Lett. {\bf 86}, 1857 (2001).
	
	\bibitem{Balatsky2013}
	D. Tanmoy and A. V. Balatsky, Nat. Commun. \textbf{4}, 1972 (2013).
	
	\bibitem{carlos} S. I. Erlingsson and  J. C. Egues, Phys. Rev. B {\bf 91},035312 (2015).
	
	
	\bibitem{Macieiko2010}
	J. Maciejko, X. Qi, A. Karch, and S. Zhang, Phys. Rev. Lett. \textbf{105}, 246809 (2010).
	
	\bibitem{nesting_1} M. D. Johannes and I. I. Mazin,  Phys. Rev. B {\bf 77}, 165135 (2008).
	
	\bibitem{nesting_2}  K. Kuroki, S. Onari, R. Arita, H. Usui, Y. Tanaka, H. Kontani, and H. Aoki
	Phys. Rev. Lett. {\bf 101}, 087004 (2008).
	
	\bibitem{nesting_3} R.  Nandkishore,	L. S. Levitov	 and  A. V. Chubukov, Nature Physics {\bf 8}, 158 (2012).
	
	\bibitem{nesting_4} D. Chowdhury and S. Sachdev, Phys. Rev. B {\bf 90}, 245136  (2014).
	
	\bibitem{Affleck1}
	I. Affleck and A.W.W. Ludwig, Nucl. Phys. B {\bf 360}, 641 (1991).
	
	\bibitem{Affleck2}
	I. Affleck,  {\it Conformal Field Theory Approach to the Kondo Effect}, Lectures notes, Acta Phys. Polon. B {\bf 26}, 1869 (1995).

	
	\bibitem{AltlandZirnbauer1997}
	A. Altland and M. R. Zirnbauer, Phys. Rev. B \textbf{55}, 1142 (1997).
	
	\bibitem{Zirnbauer1996}
	M. R. Zirnbauer, J. Math. Phys. \textbf{37}, 4986 (1996).
	
	\bibitem{Qi2008}
	X. Qi, T. L. Hughes, and S. Zhang, Phys. Rev. B \textbf{78}, 195424 (2008).
	
	\bibitem{RyuSchnyder2010}
	S. Ryu, A. P. Schnyder, A. Furusaki, and A. W. W. Ludwig, New J. Phys. \textbf{12}, 065010 (2010).	
	
	\bibitem{composite} J. Klinovaja and D. Loss, Phys. Rev. B {\bf 86}, 085408 (2012).
	
	
	\bibitem{Schliemann2003}
	J. Schliemann, J. C. Egues, and D. Loss, Phys. Rev. Lett. \textbf{90}, 146801 (2003).
	
	\bibitem{Bernevig2006}
	B. A. Bernevig, J. Orenstein, and S. Zhang, Phys. Rev. Lett \textbf{97}, 236601 (2006).
	
	\bibitem{Meier2007}
	L. Meier, G. Salis, I. Shorubalko, E. Gini, S. Sch\"on, and K. Ensslin, Nat. Phys. \textbf{3}, 650 (2007).
	
	\bibitem{Koralek2009}
	J. D. Koralek, C. P. Weber, J. Orenstein, B. A. Bernevig, S. Zhang, S. Mack, and D. D. Awschalom, Nature \textbf{458}, 610 (2009).
	
	\bibitem{Walser2012}
	M. P. Walser, C. Reichl, W. Wegscheider, and G. Salis, Nat. Phys. \textbf{8}, 757 (2012).
	
	\bibitem{MengKlinovaja2014}
	T. Meng, J. Klinovaja, and D. Loss, Phys. Rev. B \textbf{89}, 205133 (2014).
	
	\bibitem{Dettwiler2014}
	F. Dettwiler, J. Fu, S. Mack, P. J. Weigele, J. C. Egues, D. D. Awschalom, and D. M. Zumb\"uhl, arXiv:1403.3518 (2014).
	
	\bibitem{Leo2017} A. J. A. Beukman, F. K. de Vries, J. van Veen, R. Skolasinski, M. Wimmer,
	F. Qu, D. T. de Vries, B. Nguyen, W. Yi, A. A. Kiselev, M.
	Sokolich, M. J. Manfra, F. Nichele, C. M. Marcus, and L. P. Kouwenhoven, arXiv:1704.03482 (2017). 
	
	\bibitem{Burkov2011}
	A. A. Burkov and L. Balents, Phys. Rev. Lett. \textbf{107}, 127205 (2011).
	
	\bibitem{Chen2010}
	Y. L. Chen \emph{et al.}, Science \textbf{329}, 659 (2010).
	
	\bibitem{Hosseini2015}
	M. V. Hosseini and M. Askari, Phys. Rev. B \textbf{92}, 224435 (2015).
	
	\bibitem{Chang2015}
	H. Chang, J. Zhou, S. Wang, W. Shan, and D. Xiao, Phys. Rev. B \textbf{92}, 241103(R) (2015).
	
	
	\bibitem{giamarchi}
	T. Giamarchi, \emph{Quantum Physics in One Dimension} (Oxford University Press, Oxford, 2004).
	
	\bibitem{Oreg2014}
	Y. Oreg, E. Sela, and A. Stern, Phys. Rev. B \textbf{89}, 115402 (2014).
	
		\bibitem{diego} D. Rainis, L. Trifunovic, J. Klinovaja, and D. Loss, Phys. Rev. B {\bf 87}, 024515 (2013).
		
\bibitem{DiracSM3D}
	S. M. Young, S. Zaheer, J. C. Y. Teo, C. L. Kane, E. J. Mele, and A. M. Rappe, Phys. Rev. Lett. \textbf{108}, 140405 (2012).
		
	
\end{thebibliography}

\end{document}